# Preliminary results on the characterization and performances of ZBLAN fiber for infrared spectrographs

M. Iuzzolino[*a], A. Tozzi[a], N. Sanna[a], L. Zangrilli[a], E. Oliva[a]
[a]INAF – Osservatorio Astrofisico di Arcetri, L.go E. Fermi 5, 50125 Firenze, Italy (E.U.);

## ABSTRACT

Present telescopes and future extremely large telescopes make use of fiber-fed spectrographs to observe at optical and infrared wavelengths. The use of fibers largely simplifies the interfacing of the spectrograph to the telescope. At a high spectral resolution (R>50,000) the fibers can be used to achieve very high spectral accuracy.
GIANO is an infrared (0.95-2.5μm) high resolution (R=50,000) spectrometer[1] [2] [3] that was recently commissioned at the TNG telescope (La Palma). This instrument was designed and built for direct feeding from the telescope [4]. However, due to constraints imposed on the telescope interfacing during the pre-commissioning phase, it had to be positioned on the rotating building, far from the telescope focus. Therefore, a new interface to the telescope, based on IR-transmitting ZBLAN fibers with 85μm core, was developed.
In this article we report the first, preliminary results of the effects of these fibers on the quality of the recorded spectra with GIANO and with a similar spectrograph that we set-up in the laboratory. The effects can be primarily associated to modal-noise (MN) that, in GIANO, is much more evident than in optical spectrometers, because of the much longer wavelengths.

## 1 INTRODUCTION

### 1.1 General introduction to the problem of modal noise

Modal noise is a typical drawback when using optical fiber in the IR spectral range with an enough coherent source of light, and in particular in case of high performance requirements. Modes are the discrete paths light can travel in a fiber, and ideally these paths are not connected by power exchange. Actually any imperfection lead to coupling them, but this doesn't necessarily generate a burst of modal noise.

In communication fiber links modal noise appears like a signal distortion and it can be seen both in the multimode and monomode fiber. In the multimode fiber this noise is caused by selective mode loss (like a poor connector does) inside the signal, already affected by mode coupling. The result is an unpredictable signal power variation at high frequency [4]. In the monomode fiber, that includes short fiber segments, this noise appears if there is a lateral core offset at a fiber joint [7].

In high resolution spectroscopic and astronomical applications the narrow spectral field of view can be associated to an almost coherent incoming light [8]. The modal noise is explained as an add up of the interference effect at the fiber output and any mode filtering, like beam truncation. In coherent light the defects of the fiber cause asymmetry and phase shifts of the supported modes (speckle pattern), and this in addition to fiber stress and beam truncation results in modal noise [9].

The modal noise effect is also wavelength dependent. The number of modes, excited by a uniformly-illuminated fiber, is inversely proportional to $\lambda^2$, and the fewer are the modes the greater is the weight of modal noise in the overall signal recording [8]. The two mentioned fields (fiber optic communication and astronomical spectroscopy) are the most influenced by this type of noise, because of their high-level requirements about laser source quality and signal to noise ratio at high spectral resolution.

In the spectroscopic field the state of the art of the modal noise reduction techniques is the use of scrambling methods, that consist in dynamical and mechanical movements imposed at the fiber (spatial mixing of transmitted energy), or in static optical mixing of the far and near field of view (optical spatial mixing). In fiber optic communication great attention is paid to the type of connector, its quality, and cleanliness; moreover the use of an incoherent source of light can be considered, if the required communication speed allows this solution.

[*] iuzzolino@arcetri.astro.it; phone +390552752315

## 1.2 Mathematical consideration on the number of excited modes

The theoretical modeling of modal noise is currently in progress among the scientists [9], with the purpose of foreseeing this source of signal degradation and then weakening it. The overall mathematical and physical investigation is hard also because all the details about the ray tracing of light transmission inside the fiber are not yet completely clarified. Nevertheless many aspects of light's modal transmission have been deeply studied, and one of the most useful results is the formula about the number of excited modes in an optical fiber [10]. The key parameters are : the wavelength $\lambda$, the numerical aperture NA, and the diameter of the uniformly illuminated fiber core $D_{in}$, as remarked also in [11], par.3:" If one assumes that energy is equally distributed among the modes, then it can be shown (Daino, De Marchis, & Piazzola 1980) that statistical fluctuations in the number of monochromatic modes, $N_\lambda$, transmitted by the fiber introduces noise given by eq.(1), where S is the signal level and $N_\lambda$ comes from eq.(2)".

$$\text{Modal noise} = \frac{S}{\sqrt{N_\lambda}} \quad (1)$$

$$2N_\lambda = \left(\frac{\pi \cdot D_{in} \cdot NA}{\lambda}\right)^2 \quad (2)$$

Moreover it is possible to rewrite the eq. (2) in such a way to underline the dependency of the number of excited modes on the F number, and on the ratio between the uniformly illuminated fiber diameter, $D_{in}$, and the diameter of the Airy disk, $D_{airy}$, remembering that the numerical aperture NA is equal to $\frac{0.5}{F/\#}$.

$$D_{airy} = 2.44 \cdot F/\# \cdot \lambda \quad (3)$$

$$2N_\lambda = \left(1.22 \cdot \pi \frac{D_{in}}{D_{airy}}\right)^2 \quad (4)$$

## 1.3 The GIANO experience vs. the laboratory set-up

As already described above, we encountered the modal noise during the commissioning of our last instrument GIANO, the high resolution echelle spectrograph presently attached to the TNG telescope in La Palma. As we noticed an unusual and not repeatable modulation inside the stars signals, even if we used a custom mechanical scrambler, we decided to start a deeper study of the modal noise in infrared ZBLAN ($ZrF_4$-$BaF_2$-$LaF_3$-$AlF_3$-NaF) fiber, a specific case not already found in literature, to the best of our knowledge. The modal noise existence limits the radial velocity precision achievable through the spectrum measurements in the NIR range, that is the most attractive wavelength region for current planet finder studies. The spectrographic use case forced us to adopt ZBLAN fiber, because it shows a lower attenuation in the wavelength range of interest (0.9 - 2.5 micron, average attenuation of 0.18 dB/m). The main drawback of this type of fiber is its fragility (and also its softness), that causes a very little tolerance to the mechanical stresses, like small radius bending. So a possible mechanical scrambler must avoid to bend the fiber in too narrow circular paths.

As mentioned in 1.1, the fewer are the excited modes, that are involved in the light transmission, the higher is the relevance of the modal noise in the signal. GIANO works in the NIR, so that it is much more affected by this type of noise (see eq. (2)). Another challenging aspect is that, under conditions of good seeing (~0.6"), the image of the star on the fiber is 50 microns, while the fiber core diameter is 85 µm, and the decrease of image size leads to higher modal-noise (see eq. 4). Finally the optical design of GIANO includes the image slicer, a Bowen-Wallraven prism-like that cuts the spots in a half and re-aligns them so that they perfectly fit inside the spectrograph slit. This cutting phase also contributes to highlight the modal noise, although it is far to be the main noise reason.

In consideration of the above we worked to develop an experimental set-up that would be useful to simulate the modal noise, and to study it, even with the limit of lacking of some peculiarities of GIANO. We tuned the setup to achieve similar spectral resolution and level of modal noise. We discriminate between a uniform wide source (pin hole of 600 µm), comparable with GIANO flat field calibration source, and a narrower source (pin hole of 50 µm), representing the star case. The results, summarized in Table 1, show that the experimental set-up is ~2 times less affected by modal noise than GIANO. Taking into account the above considerations, it can be considered a good starting point to simulate the effects seen at the telescope using a controlled and reproducible set-up.

Detailed numerical results are listed in Table 2, Table 3.

| Light Source | Simulated fiber illumination | Typical value of laboratory modal noise | Typical value of GIANO modal noise |
|---|---|---|---|
| pin hole 600 μm | Flat Field | 0.1% rms | 0.2% rms |
| pin hole 50 μm | Star | 0.7 % rms [a] | 1.5% rms |

a: average value computed from all the test done with the small pin hole
Table 1 : Comparison of modal noise in the laboratory set-up and in GIANO

## 2  MODAL NOISE MEASUREMENTS

### 2.1  General description

The purpose of this paper is to investigate the topic of modal noise in the particular case of ZBLAN fiber, the only commercial type of fiber for applications at wavelengths longer than 2 microns. The overall scheduled work is to identify the noise, boost the causing factors, define/quantify the dependencies and attempt solutions to remove it. In this paper we focus mainly on the first step of the work, i.e. highlighting the modal noise, its dependencies, and developing tools to quantify it. In order to achieve this goal the investigation requires to build an experimental set up through which extracting the one-time dispersed spectrum of a known light source. The comparison between spectra of the same source taken at different times is used to identify and quantify the modal noise signature. In other words, we reproduce in the laboratory the canonical observation strategy: stellar spectra and flat-field are combined to correct for instrumental response.

### 2.2  Description of the experimental set-up

The experimental set up is divided in two areas: "Fiber-In" area is the first optical board that contains the light source and the optics that let the light enters in the fiber; "Fiber-out & Grating" is the second area that includes the echelle grating and the IR detector.

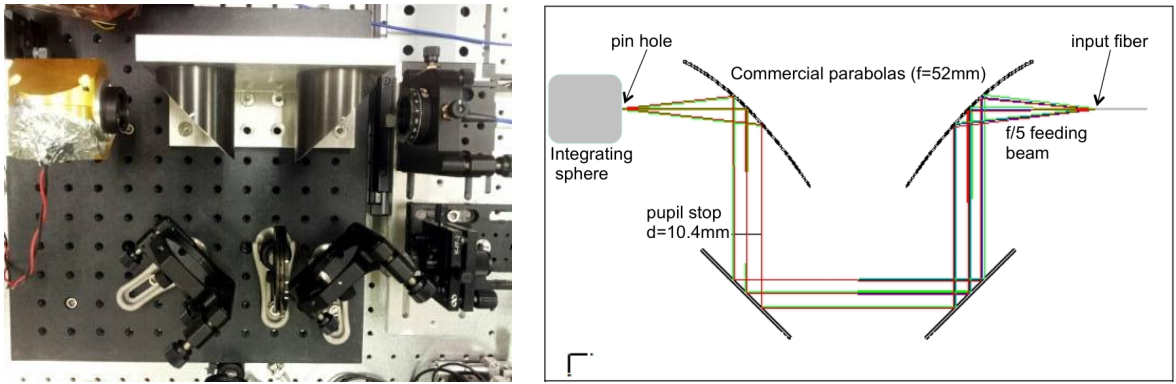

Figure 1 "Fiber-In" optical board.

The purpose of the "Fiber-In" board is to focus the input light on the entrance of the fiber. The input beam F number must be chosen according to the performance requirements of the fiber and its Focal Ratio Degradation (FRD) values. If the input F number is between 3 and 5, light travels inside the fiber with a low focal ratio degradation.

The light source is a halogen lamp inside an integrating sphere. On the exit side of the sphere there is a 50 micron pin hole (also changeable with a 600 μm pin hole). The optical set up is composed by two off-axis parabolas (commercial off-the-shelf devices with 52 mm aperture diameter and 50 mm focal length, leading to an F/# equal to 1), two flat mirrors of 52 mm diameter, and one diaphragm to change the F number. As shown in Figure 1, the beam path is the following: the diverging beam arrives on the first parabola; the collimated beam exits from the parabola, and reflects on the first flat mirror; then the beam passes through the diaphragm; the collimated beam reflects on the second flat mirror, and then it is reflected again by the second parabola, where it is focused at the entrance of the optical fiber. Due to the shapes of the flat folding mirrors and of the parabolas, the minimum F number of the beam is 2 (partial vignetting of the beam during the reflection on the first flat mirror), while the chosen working F number is 5 (1 cm aperture of the diaphragm).

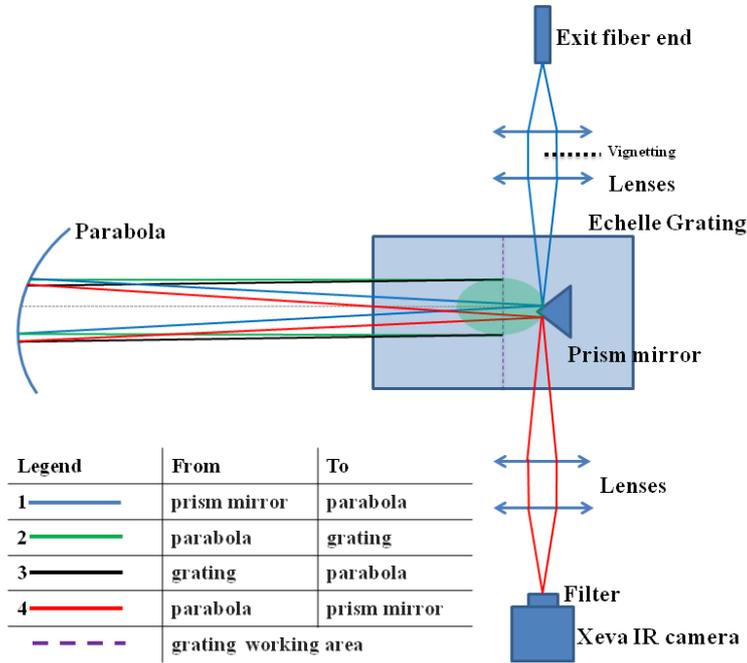

Figure 2 Sketch of the "fiber out" optical board. The lenses are commercial IR coated lenses with a nominal focal length of 100 mm (diameter 25 mm for the top ones, 50 mm for the out ones). The Parabola is a commercial one having a focal length of 4.25in diameter and 17.5in focal length.

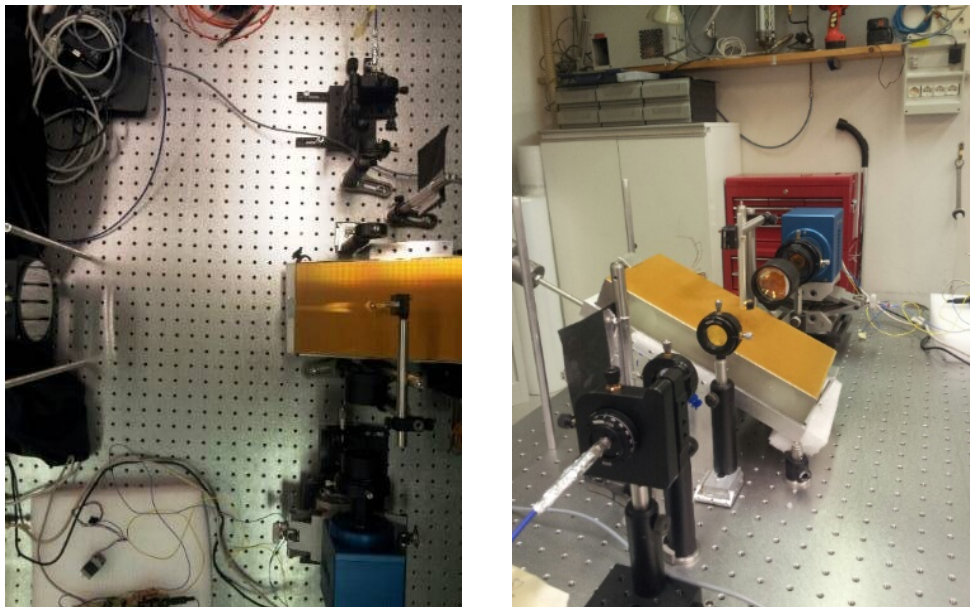

Figure 3 "Fiber-Out and grating" optical board assembled in Arcetri laboratory.

The purpose of the second board, "Fiber-out & Grating", is to use the light exiting from the fiber, in order to observe the modal noise. In this case light is dispersed as we can capture the image of the light spectrum. The study of the dispersed light allows us to understand the spectral effect of the modal noise, and thanks to a narrow wavelength selection, the spatial effect is also visible (no spectral orders overlapping).

As shown in the sketch of Figure 2 and in Figure 3 top view, the light path is the following: the board input light is the exiting light from the fiber end; then optics rearrange the beam and direct it toward the grating; finally the beam is focused again on the IR detector. The optics are the lenses that focus the input beam on one side of a roof prism mirror, suspended over the grating. The focused beam is reflected by the roof prism and arrives on the wide parabola. This last element reflects and collimates the beam, that then impacts on the grating. The dispersed beam is then reflected by the grating toward the parabola again. The parabola focuses the beam on the other side of the roof prism mirror. That side of the mirror reflects the beam toward the exit lenses in front of the detector. The exit lenses collimate and focus the beam on the detector.

The used ZBLAN multimode fiber has a core diameter of 85 micron. The dispersing element is an echelle grating, with 24.2 lines/mm and 63° blaze angle. A narrow band filter in front of the detector selects the grating order centered at 1645 nm (with a bandwidth of about 8 nm). The camera model is the Xenics Xeva-796 (XC131), based on an InGaSb detector with a 30x30 micron pixel size, 320x256 pixel image sensor, with 0.9 - 1.7 micron wavelength response. The measured conversion factor is approximately equal to 20 $e^-$/ADU. The wavelength scale on the recorded spectra is about 0.03 nm/pixel. The wide parabola has a diameter of 114 mm, while the output lenses have a diameter of 52 mm. The resolving power of the spectrograph illuminated by a 85 micron fiber is about R~20,000.

## 2.3 General description of the measurements

The general concept of the experiment is to acquire many frames so that the nominal signal to noise ratio, defined by the read-out and photon noises, is sufficiently high to highlight the modal noise. The analysis is done on the mean value computed on a set of frames (400, 100, 50, or 25, according to the type of test).

The experimental set-up is our accuracy limiting element, and this constrain imposed us to find the best data processing to pointedly see the modal noise. We experienced that the modal noise can be seen in two different ways of analysis. The first approach consists in the extraction of the spectrum from one mean frame, and then in the comparison between this spectrum and the others coming from the subsequent mean frames. The second approach consists in the direct subtraction of one mean frame with the subsequent. Each frame is in fact the image of the dispersed light in the considered wavelength range, and this comparison highlights strange and irregular modulation in the signal, visible even by eye. This second type of study is faster, and is especially well suited for quick modal noise evaluation during the laboratory activity.

In order to carry out a larger number of tests, to automate the acquisitions, and to take each frame at the same time distance and with the same exposure time, a triggering configuration has been implemented on the camera, so that the running of the process is manageable from one IDL custom procedure.

The tests we did are: preliminary tests, to choose the best data handling and the best experimental set-up starting condition, and dependencies tests, to understand the modal noise response with varying parameters.

## 2.4 Preliminary tests

The data handling question led us to try two different solutions. The first option was developed as follows: acquisition of light frames (group of 100 images, exposure time equal to 500 ms for each frame, halogen lamp with 4 Ampere current as light source, to achieve a peak signal of 1500 ADU/pixel) and dark frames; computation of the mean frames and mean dark frames; subtraction of the two mean frames and extraction of the spectrum (IRAF procedure). Repeating these steps several times, we get spectra that can be combined and subtracted one with the other to show their relative differences.

The second tested option was a faster and easier way to underline the modal noise. It consisted in the following steps: acquisition of light frames (in the same condition as before); computation of the mean frames; subtraction of the first two subsequent mean frames. This last simple action showed the bi-dimensional (2D) dispersed spectrum image with a morphological modulated aspect, so that the 2D spectrum shows a twisted pattern along the vertical direction (the spectral direction) and fewer also horizontally (the spatial direction). This original aspect suggested to name the 2D image of the dispersed spectrum as "scooby-doo image". We repeated these steps several times and we appreciated the visibility of this twisted signal; the image subtraction together with the controlled experimental set up lead us to the deduction that this effect was due to the modal noise.

In order to search for a static configuration that would enhance the visibility of this noise, we changed the fiber arrangement all along its length, starting from a rolled up position, then trying a vertically hooked path, so that the fiber was more extended, and finally a horizontal wide curve path. The more the path was stretched out, the more the modal noise was evident. The avoidance of narrow bending (less than 30 cm radius) also contributed to the highlight of the modal noise. The last tried configuration was the most modal noise affected, but further following tests would have shown that the static position of the fiber has a temporary influence on the modal noise visibility, and this influence decreases with the passage of time. The final chosen fiber position was a horizontal path on the optical bench with about 4 wide curves.

## 2.5 The modal noise evaluation tool

Before proceeding to the second type of tests, it is worth to describe the tool we used to "quantify" the presence of the modal noise. Through the IDL software, we developed a procedure that is able to extract the mono-dimensional (1D) spectrum from the "scooby-doo" image, to remove any baseline in the signal, and to compute the FFT and the power spectrum of the 1D spectrum. Within this procedure, a preliminary quantitative estimation of modal noise inside the 1D spectrum was done thanks to three representative parameters. The intensity is identified by the Root Mean Square of the 1D spectrum, it is in fact suggestive of the oscillation amplitude of the signal. The morphological regularity of the "scooby-doo" image and of the 1D spectrum features is shown by the power spectrum function, and the linked evaluation parameter is the integral of the power spectrum within the frequency range of interest (from 0.0078 cy/pix to 0.0508 cy/pix).

The 1D spectrum is the function that reveals the modal noise signature, with an evident irregular signal modulation, so it is the most pertinent estimable element to be monitored. As described above, the "scooby-doo" image is obtained as the difference between two subsequent mean frames. The spectrum computation consisted in: identifying the number of columns in the 2D "scooby" image that should include the signal of interest; summing the selected columns; normalizing the spectrum, that is dividing the obtained 1D signal by a reference parameter. This last parameter is the mean value of the signal computed as the sum of the same numbers of columns in one of the two considered mean frames. This last step implies that the RMS is later expressed as a percentage of the mentioned reference parameter.

## 2.6 Experimental preliminary results

The more relevant and intense phase of our study was to try to understand under which circumstances the appearance of modal noise is promoted, taking into account that our main goal in this experimental set-up is to verify the possibility to simulate in laboratory the behavior of GIANO-TNG and from these investigations we would like to have a helpful tip for the correct use of fiber-fed prelist systems in IR spectrographs.

Firstly we identified a set of parameters that could have influence on the extent of the modal noise, or that in anyway could be linked with such type of noise. Then we performed the appropriate tests to get the answer: in Table 2, Table 3 and Table 4 the preliminary experimental results are reported.

The identified parameters to simulate the on-sky observing modalities are the followings: the size of the pin hole (50 μm or 600 μm); the mechanical scrambling of a portion of the fiber vs. a static position; the influence of the defocus, intended as a way to simulate a time variable fiber lighting; the influence of vignetting of the light at the spectrometer entrance (none or 50%); the far field variation with respect to the modal noise appearance; the variation of the F number.

According to the literature [9], the influence of vignetting consists in increasing the modal noise evidence. Indeed, both tests with 50 μm pin-hole and 600 μm pin-hole, with a scrambling motion or in a static condition, showed that the mean value of the RMS in case of 50% of pupil vignetting is almost double than the value with no vignetting.

The pin-hole influence reveals that in case of smaller dimension (50 μm) the rms is higher (0.7%) than in the case of wider pin hole (about 0.1%). In the GIANO-TNG astronomical-spectroscopic instrument the first case is comparable to the observation of a star, for which the angular extension on the fiber-in is smaller than the diameter of the fiber, while the second case simulates the observation of a uniform source as in the reference light flat field exposure. Moreover we observed that with the smaller pin hole the spectrum is also strongly influenced by the illumination accuracy from the external source, as showed by the RMS value in case of defocus. With the 600 μm pin hole, much larger than the fiber core diameter (85 μm), the defocus influence is of course not relevant. Moreover with the bigger pin hole the integral spectral parameter is less sensitive to the vignetting.

Due to the modal noise visibility resolution in our experimental set-up, we performed every activity that would enhance the presence of this type of noise. In all the tests, with exposure time of 500 ms and different numbers of acquired frames, it was noticed that a low frequency (1 Hz) scrambling or impulsive and short movements act as a perturbation of the fiber modes arrangement, and with a short enough exposure time it enhances the evidence of modal noise. Moreover, after a low frequency scrambling activity a relaxing time is needed in order to have the modal noise signature inside the spectrum disappear. On the other hand a longer exposure together with higher frequency scrambling activity is helpful in attenuating this noise, and the scrambling activity mentioned in the following refers to this last type. The indication of "impulsive shaking" and "no-like" as a scrambling status indicates the low frequency perturbation activity or impulsive perturbation mentioned above.

|   | **WIDE PIN HOLE measurements** | | | |
|---|---|---|---|---|
|   | Pin hole | Exposure time | Max signal | No Vignetting |
|   | 600 µm | 500ms | 1500 counts | |
|   | **Test Results** | | | |
|   | Scrambling | n° images | focus vs. defocus | RMS |
| 1 | yes | 400 | in focus | 0.097% |
| 2 | yes | 800 | in focus | 0.111% |
| 3 | yes | 300 | in focus | 0.064% |
| 4 | yes | 400 | in focus | 0.105% |
| 5 | yes | 400 | defocus | 0.079% |
| 6 | yes | 400 | defocus | 0.160% |
|   | *Mean value of RMS* | | | **0.103%** |

Table 2 : Example of test using a wide pin-hole to simulate a uniform illumination of the fiber core (85 µm).

An interesting result comes from the defocus tests with the small pin hole. They consisted in the subtraction of two mean frames, the first was taken in the extra-focal position, and the second was acquired in the intra-focal position. The effect on the 1D spectrum has a modal noise like - signature with a consisted value of RMS (0.6%), and the mechanical scrambling showed an attenuating effect.
The far field test showed that its appearance is totally independent from: the size of the pin hole, the lighting accuracy at the input fiber end, and the mechanical scrambling.
Last tests concerned the investigation of different optical choices for the illumination of the input fiber end. The tests concerned two options: in the first case the illumination of the input fiber end was done in the focal plane ("in-focus fiber illumination"); in the second case the illumination of the input fiber end was done in the pupil plane ("in-pupil fiber illumination"). The experimental results showed that no significant difference in terms of RMS of the extracted spectrum occurred between the in-focus illumination and the in-pupil illumination of the input fiber end.
The following tables summarize the test characteristics and the average results for repeated measurement

|   | **SMALL PIN HOLE measurements** | | | |
|---|---|---|---|---|
|   | Pin hole | Exposure time | Number of images | Max signal | No Vignetting |
|   | 50 µm | 500 ms | 400 | 1500 counts | |
|   | **Test Results** | | | |
|   | *Test Defocus @ not shaking* | | | |
|   | Scrambling | F/# | focus vs. defocus | RMS |
| 1 | no | 5 | defocus | 0.63% |

| | | | | | |
|---|---|---|---|---|---|
| 2 | no | 5 | defocus | | 0.66% |
| 3 | no | 5 | defocus | | 0.63% |
| | *Mean value of RMS* | | | | **0.64%** |
| | ***Test Defocus @ shaking*** | | | | |
| 4 | yes | 5 | defocus | | 0.50% |
| 5 | yes | 5 | defocus | | 0.32% |
| 6 | yes | 5 | defocus | | 0.28% |
| | *Mean value of RMS* | | | | **0.37%** |
| | ***Test InFocus @ shaking*** | | | | |
| 7 | yes | 5 | in focus | | 0.25% |
| 8 | yes | 5 | in focus | | 0.23% |
| 9 | yes | 5 | in focus | | 0.22% |
| | *Mean value of RMS* | | | | **0.23%** |
| | ***Test InFocus @ impulsive shaking*** | | | | |
| 10 | no-like | 5 | in focus | | 0.94% |
| 11 | no-like | 5 | in focus | | 0.83% |
| 12 | no-like | 5 | in focus | | 0.82% |
| | *Mean value of RMS* | | | | **0.87%** |
| | ***Test Fnum @ impulsive shaking*** | | | | |
| 1 | no-like | 10 | in focus | | 1.34% |
| 2 | no-like | 10 | in focus | | 1.29% |
| | *Mean value of RMS* | | | | **1.31%** |
| | ***Test Fnum @ shaking*** | | | | |
| 3 | yes | 10 | defocus | | 0.24% |
| 4 | yes | 10 | defocus | | 0.23% |
| 5 | yes | 10 | defocus | | 0.14% |
| | *Mean value of RMS* | | | | **0.21%** |

Table 3 : Example of measurements performed with a small pin-hole (50 µm image diameter) to illuminate the fiber core (85 µm).

| Test Case name | Test Experimental Condition | | | | | | Results |
|---|---|---|---|---|---|---|---|
| | pin hole 50 µm | pin hole 600 µm | Scrambling | Impulsive scrambling | Input Defocus | Input F/# | RMS |
| *Wide pin-hole test* | | X | X | | | 5 | **0.10%** |
| | | X | X | | X | 5 | |
| *Test Defocus@NotShaking* | X | | | | X | 5 | **0.64%** |
| *Test Defocus@Shaking* | X | | X | | X | 5 | **0.37%** |
| *Test InFocus@Shaking* | X | | X | | | 5 | **0.23%** |
| *Test InFocus@Impulsive shaking* | X | | | X | | 5 | **0.87%** |
| *Test Fnum@Impulsive shaking* | X | | | X | | 10 | **1.31%** |
| *Test Fnum@Shaking* | X | | X | | | 10 | **0.21%** |

Table 4 : Tests summary result table. Fiber core diameter is 85 µm.

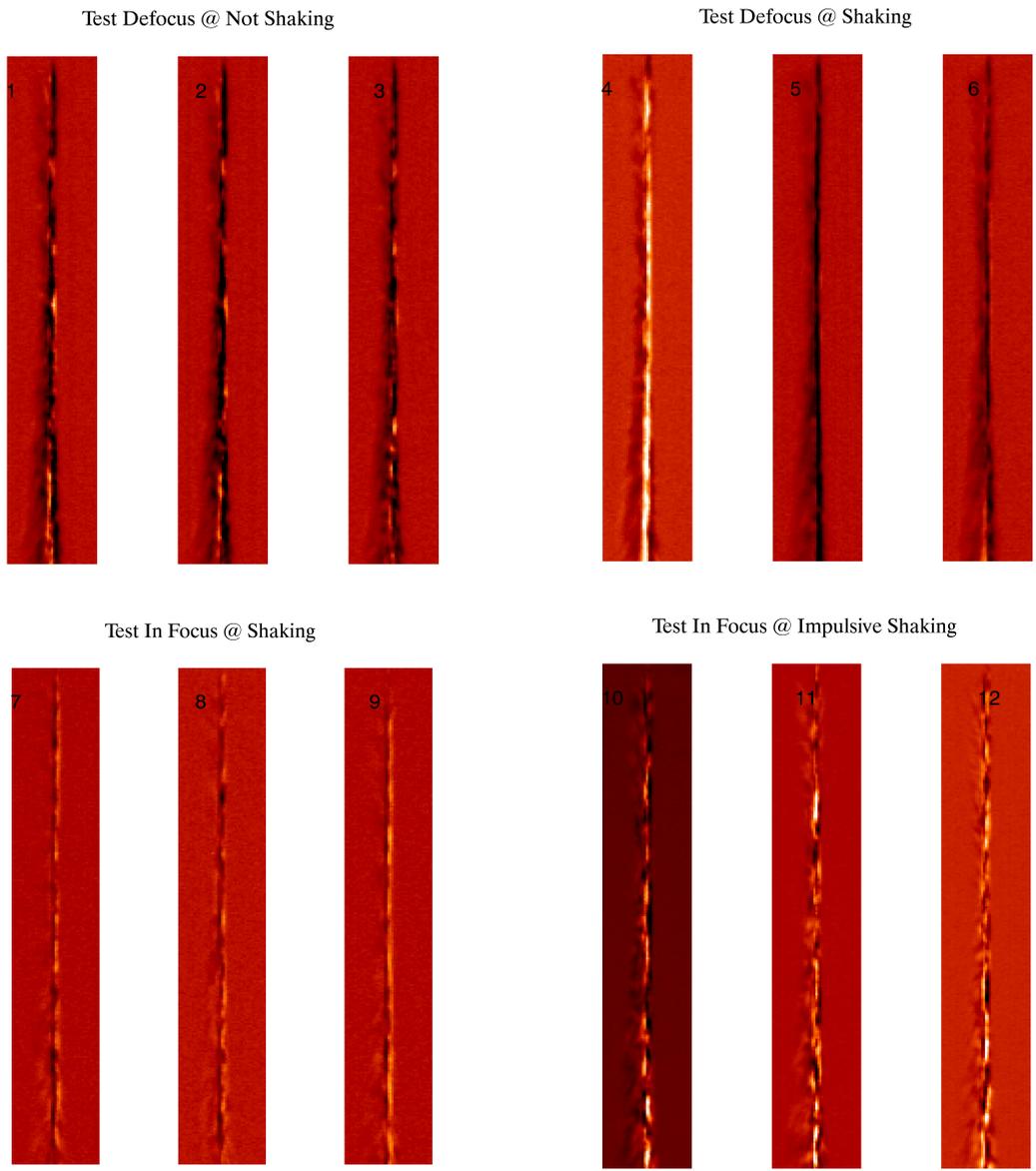

Figure 4  Example of raw data (scooby-doo images) as acquired by the XEVA IR camera in the wavelength region of 1640 nm. The vertical axis is the dispersion one: the whole spectral height is 8 nm. The fiber out dimension on the XEVA detector plane is 3 pixel. Each images correspond to the relative test case, as numbered in Table 3

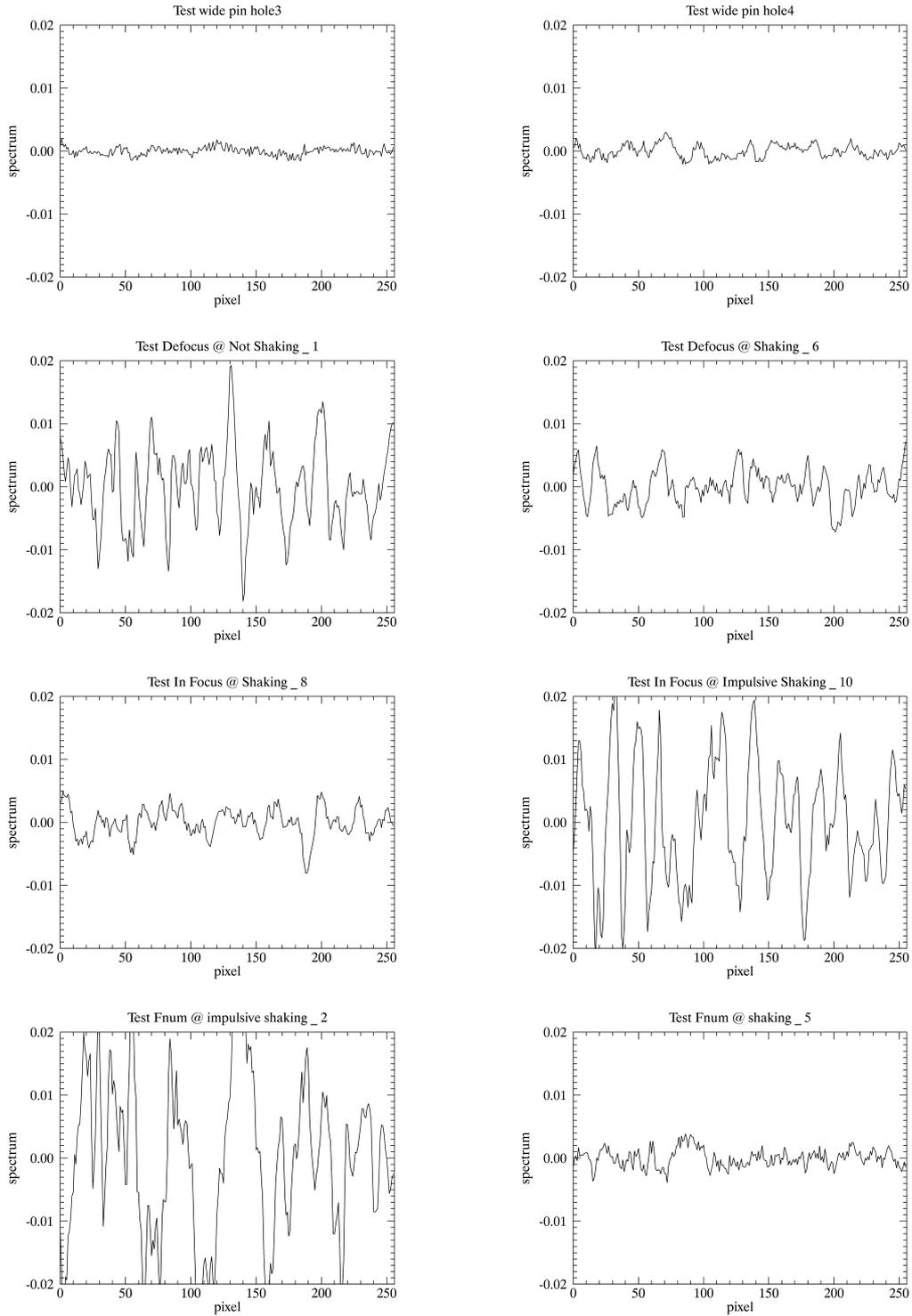

Figure 5 : Example of 1D spectrum, extracted by the corresponding scooby-doo images of Figure 4, and related to the numbered test case as reported in Table 2 and in Table 3

## 3   CONCLUSIONS

This first phase of our investigation was the development of an experimental set up that highlights the modal noise and its dependencies. As already described, the preliminary tests were necessary to define a convenient way to manage data in order to visually identify the modal noise signature. The dependencies tests showed the relation between the worst drawback we have found in the IR fiber, that is the modal noise, and the main system variables that characterize an optical system. During this investigation, a different source of noise was discovered, that is due to time variable fiber illumination (in this case a defocus), and further studies will be essential, because of its possible effects on astronomical observations.

The next step was a deeper analysis of the modal noise characteristics, and tests to identify the methods to attenuate that kind of noise. The final goal of our work is finding a solution for modal noise in this particular case of ZBLAN fiber, and in general in the IR wavelength range.

## 4   ACKNOWLEDGEMENT

This work has been developed thanks to the INAF financial support through the grants "TECNO-2011" and "TREX-2011".